\documentclass [10pt]{article}
\usepackage[T1]{fontenc}
\usepackage[final]{epsfig}
\usepackage{amssymb}
\usepackage{multicol}
\usepackage{graphics}
\usepackage{color}
\usepackage{ntimes}

\newcommand{\BE}{\begin{equation}}
\newcommand{\EE}{\end{equation}}

\begin{document}

\title{Parking and the visual perception of space}
\author{\textbf{Petr \v Seba${}^{1,2,3}$}\\
\small{${}^1$ University of Hradec Kr\'alov\'e, Hradec Kr\'alov\'e
- Czech Republic}\\
\small{${}^2$ Institute of Physics, Academy of Sciences of the
Czech Republic, Prague - Czech Republic}\\
\small{${}^3$ Doppler Institute for Mathematical
Physics and Applied Mathematics,}\\ \small{Faculty of Nuclear
Sciences and Physical Engineering,
Czech Technical University, Prague - Czech Republic}}

\normalsize

\maketitle

\begin{abstract}
Using measured data we demonstrate that there is  an amazing
correspondence among the statistical properties of spacings between
parked cars and the distances between birds perching on a power
line. We show that this observation is easily explained by the fact
that birds and human use the same mechanism of distance estimation.
We give a simple mathematical model of this phenomenon and prove its
validity using measured data.
\end{abstract}

Everyone knows that to park a car in the city center is problematic.
The amount of the available places is limited and it has to be
shared between too many interested parties.  Birds face the same
problem when a flock tries to perch on an electric line. The common
problem (we will call it "the parking problem" in the following
text) is to place oneself safely between the two parties limiting
the available space from the left and from the right. This leads
necessarily to an interaction preventing collisions between the
participants and making the parking maneuver not random. It should
be therefore not a surprise that the "Random car parking model"
introduced by Renyi \cite{Renyi} (see also \cite{evans} and
\cite{ca} for review) does not  describe the real parking data
correctly \cite{rawal}, \cite{seba}. The exact character of the
interaction is unknown and hardly describable in the physical terms.
It is however clear that it is primarily triggered by the brain and
then mediated through the muscles (by pressing the accelerator/brake
pedal (cars) or by flopping the wings (birds)). Though the
particular locomotive activity is different the basic neural
regulatory mechanisms can be similar in both cases. The complex
interplay between the individuals participating on the parking
process leaves footprints in the statistics of the nett distances
(spacings) between the neighbors. (Nett distance means the distance
between the front and tail bumpers of the two adjoined parked cars
or between the wings of two birds perching side-by-side.) This
spacings can be measured. The amazing result is that obtained
spacing distribution does not depend on the kind of the parking
objects - it seems to be the same for the cars as well as for the
birds.\\

To be more specific: the subjects of our study are cars parked in
parallel in the city center and starling flocks perching on the
electric power lines. (For the dynamics of a starling flock see
\cite{starling} and \cite{starling2}). Our aim is to present a
simple theory that combines the psychophysical knowledge with a
simple mathematical model and is able to explain the observed
universality of the spacing statistics. The psychophysical part  is
based on the visual perception of space. The perception mechanism is
very old and shared by many species ranging from birds to mammals.
For man it is processed automatically and without a conscious
control. We will use it to understand the statistical properties of
the distances between the neighboring competitors in a situation
when the available space is limited. To illustrate the approach we
focus on the spacing distribution (the bumper to bumper distances)
between the cars parked in parallel. We will assume that the street
segment  used for parking has a length $L$ and is free of any kind
of  obstructions. The drivers are free to park the car anywhere
provided they find an empty space to do it. We suppose also for
simplicity that all cars have the same length $l_0$. Since many cars
are cruising for parking there are not free parking lots and a car
can park only when another parked car leaves. To simplify the
formulation of the problem and to avoid troubling with the boundary
effects we assume that the street is very long: $L>>l_0$. To park a
car of a length $l_0$ one needs (due to the parking maneuver) a lot
of a length larger then approximately $1.2 l_0$. So on a street of
the length $L$ the number of the parked cars equals to
$N\approx[L/(1.2 l_0)]$. Denoting by $D_k$ the spacing between the
car $k$ and $k+1$ we get
\begin{equation}
\sum_{k=1}^{N}D_k = L - N l_0.
\label{total}
\end{equation}
and the spacings $D_k$ are hence not statistically independent. But
for a street that is long enough this constrain does not play any
role and we will treat $D_k$ as being independent. Since the parked
cars gradually leave the street and are replaced by new cars parking
into the vacant lots the distances $D_k$ undergo continual changes.
The spacing distribution is obtained as a steady solution of this
process. Suppose that in one time step only one car can leave the
street. (We intentionally omit the situations when two and more
neighboring cars leave simultaneously.) The related distance mapping
goes as follows: In the first step one randomly chosen car leaves
the street and the two adjoining lots merge into a single one. In
the second step a new car parks into this empty space and splits it
again into two smaller lots. The splitting is random with certain
preference reflecting the parking maneuver. When a car leaves the
two neighboring spacings - say the spacings $D_n,D_{n+1}$ - merge
into a single lot of a length $D$:
\begin{equation}\label{cdistances}
    D=D_n+D_{n+1}+l_0.
\end{equation}
A new parked car splits $D$  again into two spacings
 $\tilde D_n, \tilde D_{n+1}$:
\begin{eqnarray}
 \nonumber
  \tilde D_n &=& a(D-l_0) \\
  \tilde D_{n+1} &=& (1-a)(D-l_0).
  \label{cnewdistances}
\end{eqnarray}
where $a\in (0,1)$ is a random variable with a probability density
$q(a)$. The distribution $q(a)$ describes the parking preference of
the driver. We assume that all drivers have the same habits, i.e.
they share the same $q(a)$. The meaning of the variable $a$ is
straightforward. For $a=0$ the car parks immediately in front of the
car delimiting the parking lot from the left without leaving any
empty space (an very unpleasant way to park a car) . For $a=1/2$ it
parks exactly to the center of the lot $D$ and for $a=1$ it stops
exactly behind the car on the right. Combining (\ref{cdistances})
and (\ref{cnewdistances}) gives finally the distance mapping:
\begin{eqnarray}
 \nonumber
  \tilde D_n &=& a(D_n+D_{n+1}) \\
  \tilde D_{n+1} &=& (1-a)(D_n+D_{n+1}).
  \label{cnewdistances2}
\end{eqnarray}
and the car length $l_0$ drops out.\\

For various choices of $n$ the mappings (\ref{cnewdistances2}) are
regarded as statistically independent . Since all the cars are equal
and all the drivers have the same parking habits the joint distance
probability density $P(D_1,...,D_{N})$ has to be exchangeable ( i.e.
invariant under the permutation of the variables) and invariant with
respect to (\ref{cnewdistances2}). Its marginals $p_k(D_k)$  (the
probability density of a particular spacing $D_k$) are identical:
\begin{equation}\label{distrib}
    p_k(D_k)=p(D_k)=\int_{D_1+..+D_N=L-N l_0} P(D_1,...,D_{N})
    dD_1..dD_{k-1}dD_{k+1}..dD_{N}.
\end{equation}
We suppose that the parking maneuver is known and described by the
distribution $q(a)$. For simplicity we assume a symmetric maneuver,
i.e. $q(a)=q(1-a)$ with a maximum at $a=1/2$.  The symmetry of
$q(a)$ means that the drivers are
 not biased to park more closely to a car adjacent from the behind
or from the front. For the given $q(a)$ we look for a distribution
of $D_n$ that is invariant under the transform
(\ref{cnewdistances2}). In other words we try to solve the equation
\begin{equation}\label{perp}
     D \triangleq a(D+D')
\end{equation}
where $D'$ is a statistically independent copy of the variable $D$
and the symbol $\triangleq$ means that the left and right hand sides
of (\ref{perp}) have identical statistical properties.
Distributional equations of this type are mathematically well
studied - see for instance \cite{dev} - although not much is known
about their exact solutions. In particular it is known that for a
given distribution $q(a)$  the equation (\ref{perp}) has an \it
unique solution \rm that can be obtained numerically. We are however
preferably interested in an explicit result. This can be obtained
when we choose a suitable distribution $q(a)$. Fortunately the class
of the standard $\beta$ distributions fulfills this requirement.
This is a direct consequence of the following statement \cite{duf1}:
\\
\\
\textbf{Statement:} Let $D_1,D_2$ and $a$ be independent random
variables with the distributions: $D_1\sim \Gamma(a_1,1)$, $D_2\sim
\Gamma(a_2,1)$ and $a\sim \beta(a_1,a_2)$. Then $a(D_1+D_2)\sim
\Gamma(a_1,1)$.
\\
\\
The symbol $\sim$ means that the related random variable has the
specified probability density.  $\Gamma(g,1), \beta(g_1,g_2)$ denote
the standard gamma and beta distributions respectively. For a
symmetric parking maneuver we have to take $g_1=g_2=g$. The
variables $D_1,D_2$ are then equally distributed and the solution of
(\ref{perp}) reads $D\sim\Gamma(g,1)$.  For the moment $g$ is a free
parameter. We will show, however, that psychophysical arguments (the
visual controlling mechanism) set this parameter to  $g=3$. The
point is that for small $a$ the behavior of $q(a)$ reflects the
capability of the driver/bird to estimate small distances and to
avoid collisions during the ranging process.\\

A distance perception is a complex task. For human  there are
several cues for it. Some of them are monocular (linear perspective,
monocular movement parallax etc.), others oculomotor (accommodation
convergence) and finally binocular (i.e. based on the stereopsis).
All of them work simultaneously and are reliable under different
conditions - see \cite{jacobs} for more details. For the ranging
however the crucial information is not the distance itself but a
combination of the distance and the approach velocity. To control
the situation the estimated time to collide with the neighbor play a
predominant role. Even more: a faultless avoidance of the collisions
with the surrounding objects is important for the survival of the
given species. This is why a special mechanism evaluating the
collision time
 has been developed early in the evolution.
This mechanism is old and shared with the majority of the species
including birds and mammals. It has been argued in a seminal paper
by Lee \cite{lee} that the estimated time to collision is
psychophysically derived using a quantity defined as the inverse of
the relative rate of the expansion of a retinal image of the moving
object (this rate is traditionally denoted as $\tau$). Behavioral
experiments have indicated that $\tau$ is indeed controlling actions
like contacting surfaces by  birds and mammals (including man) - see
\cite{weel},\cite{hop},\cite{shra}. Moreover the studies have
provided abundant evidence that $\tau$ is processed by a specialized
neural mechanisms in the the retina and in the brain \cite{far}. The
approved hypothesis is that $\tau$ is the informative variable for
the collision free motion - see \cite{fajen} for review.\\

Denote by $\theta$  the instantaneous angular size of the observed
obstacle (for instance the front of the car we are backing to during
the parking maneuver). Then the estimated time to contact is given
by

\begin{equation}\label{tau}
     \tau=\frac{\theta}{d\theta/dt}
\end{equation}
(Note that $\tau$ gives the contact time without explicitly knowing
the mutual velocity, the size of the object and its distance.) Since
$\theta (t) =2 \arctan(L_0/2D(t))$ with $L_0$ being the width of the
obstacle and $D(t)$ its instantaneous distance, we get
\begin{equation}\label{tau2}
    \tau(t)=-\frac{L_0^2+4D(t)^2}{2L_0 (dD(t)/dt)}
    \arctan\left(\frac{L_0}{2D(t)}\right).
\end{equation}
For $D>>L_0$ and a constant approach speed $v=-dD/dt$ the estimated
time to contact simply equals to the physical arrival time:
$\tau=D/v.$ For small distances, however, $\tau\approx D^2/(v L_0 )$
and the estimated time to contact decreases \it quadratically \rm
with the distance. We assume that the endeavor to exploit small
distances is \it directly proportional to the estimated time to
contact \rm. This means in particular that if $\tau$ evaluated in
the course of the approach is small (i.e. a collision is impending)
the further proximity is stopped. In other words: the probability
that a driver will drive up very close to an obstacle is
proportional to the corresponding $\tau$. Based on this principle we
get  $q(a)\approx a^2$ for small $a$. Since $q(a) = \beta(a,g,g)$
this sets $g=3$ and the normalized spacing distribution  reads
simply
\begin{equation}\label{clear}
p(D)=\Gamma(D,3,1/3)=\frac{27}{2} x^2 e^{-3 x}
\end{equation}
(The mean spacing equals to 1.)\\

\begin{figure}
\begin{center}
  \includegraphics[height=9cm,width=15cm]{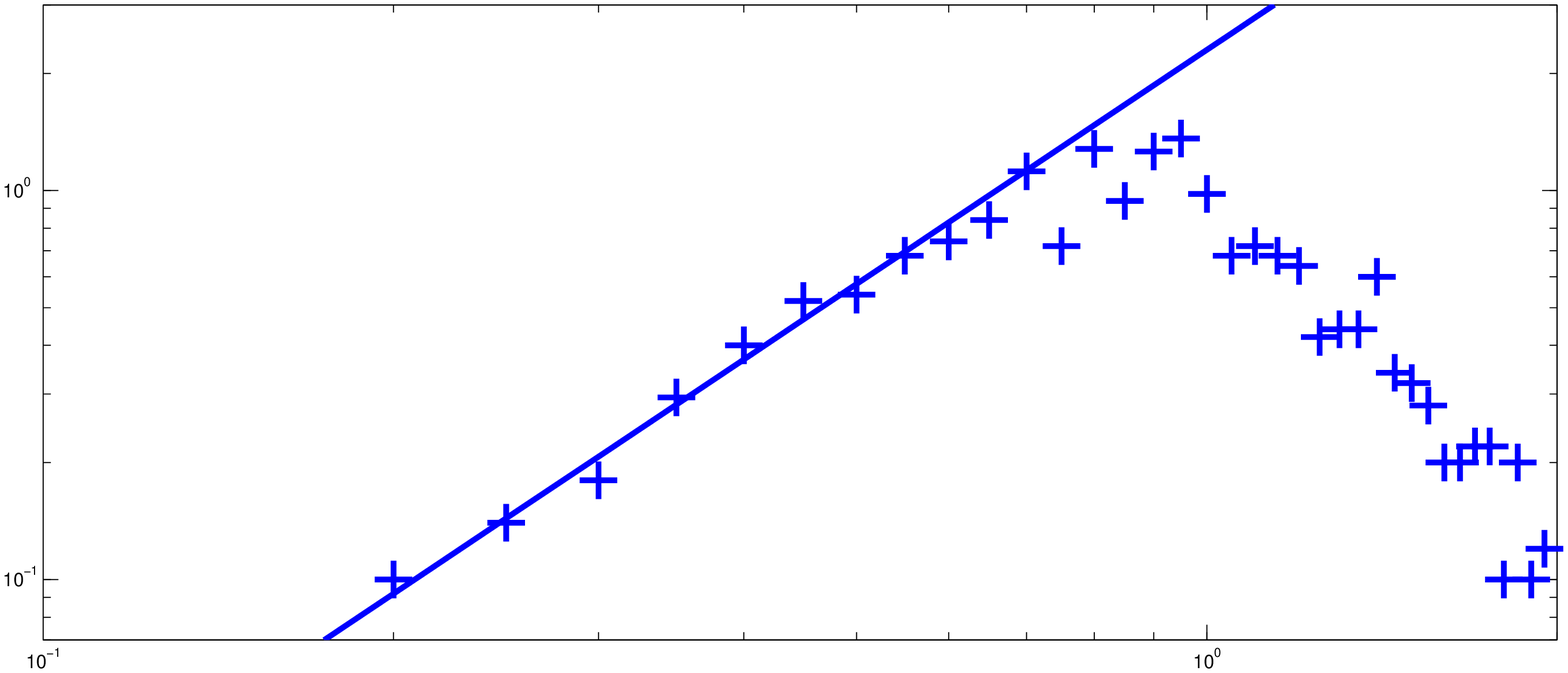}\\
\end{center}
  \caption{
  The probability density $p(D)$ evaluated for the measured data
  (crosses)
  is compared with the function $2.3*D^2$ (full line). The agreement for small $D$
  is evident.A log-log scale is used.
  }
  \label{tau3}
\end{figure}

The mechanism works so to say in the background, i.e. without being
conscious. In addition it is identical by man and by birds. This
fact has a measurable consequence: the spacing distribution for cars
parked in parallel and for birds perching on a power line should be
the same. And exactly this will be now demonstrated with the help of
the measured data. Let us first check the validity of the hypothesis
predicting that the small distance behavior of the drivers is $\tau
-$controlled. There exist a simple observation that enables us to do
it: cars stopping on a crossing equipped with the traffic lights. If
the light is red the cars stop and form a queue. We assume that the
drivers stop independently and in a distance to the preceding car
that is controlled  by $\tau$. So the clearance statistics should
give an evidence of the validity of the $\tau -$ hypothesis. In
particular the clearance probability density  should behave
quadratically on small distances. (Note that there is also an direct
experiment measuring the clearance statistics in laboratory
conditions - see \cite{gadgil},\cite{GREEN}.) We photographed the
car queues on a crossing in a front of the stoplight. The
photographes were taken all from one spot and at the same daytime.
The clearances were finally obtained by a photograph digitalization.
Altogether we extracted 1000 car distances from one particular
crossing in the city of Hradec Kralove (Czech republic) and
evaluated the corresponding probability density $p(D)$. Similar
measurement has been done also on several crossings in Prague and in
Beijing - see \cite{krbalek},\cite{li}. There is not a distance
updating inside the queue  - once stopped the driver waits for the
green light. So the obtained distance density $p(D)$ is just
proportional to $\tau$: $p(D)\approx\tau\approx D^2$ for small $D$.
The result of our measurement is plotted on the Figure \ref{tau3}
and illustrates a
nice confirmation of the $\tau-$hypothesis.\\

\begin{figure}
\begin{center}
  \includegraphics[height=9cm,width=15cm]{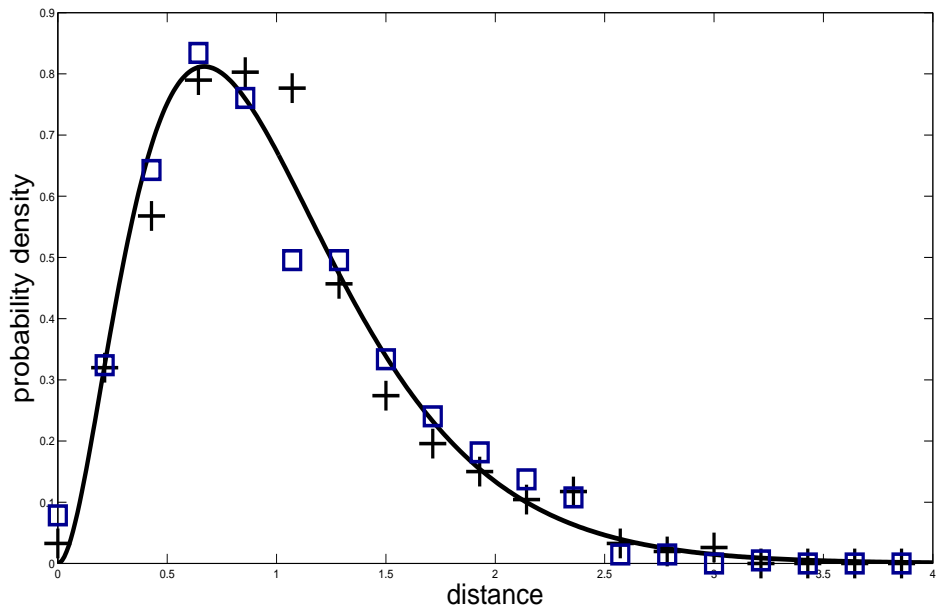}\\
\end{center}
  \caption{The probability density of the distances between the parked cars
  (crosses)
  and perching starlings (squares) is compared with the prediction of the
  theory (full line). The mean distance is normalized to 1.
  }
  \label{birds}
\end{figure}

Let us now come to the experimentally measured clearance
distribution obtained for cars parked in parallel and for birds
perching on the power line. In both cases the "parking segment" was
full, i.e. there was  not a free space for an additional
participant. In addition the segments under consideration were long
and contained a large number of objects. So $N>>1$, and the
constrain (\ref{total}) does not play a substantial role. So the
resulting distribution is invariant under the transform
(\ref{cnewdistances2}) and  the solvable variant of the model with
$q(a)\sim\beta(3,3)$ gives $p(D)=\Gamma(D,3,1/3)$. To verify the
prediction we measured the bumper to bumper distances between the
cars parked in the center of Hradec Kralove (Czech Republic). The
street was located in a place with large parking demand and usually
without any free parking lots. Moreover it  was free of any dividing
elements, side ways and so on. Altogether we measured 700 spacings
under this conditions. For the birds we photographed flocks of
starlings resting on the power line during their flight to the
south. The line was "full" and other starlings from the flock were
forced  to use the space vacated by another starling or to use
another line to perch. The bird-to-bird distances were obtained by a
simple photograph digitalization - altogether 1000 spacings. After
scaling both data sets to the mean distance equal to 1 the results
were plotted on the Figure \ref{birds} and compared with the
prediction (\ref{clear}).
The probability distributions resulting from these data seems to be
(up to the statistical fluctuations) identical  and  in a good
agreement with the model prediction. This fact is amazing since the
used "hardware" is fully different. The underlying psychophysical
mechanism, is, however, identical. (For the experimental results
concerning the relevance of $\tau$ for the space perception of
pigeons see \cite{pigeon}.)\\
\\
To summarize we have demonstrated that the clearance statistics
between parked cars and perching birds is very similar. This
surprising observation can be understood as a consequence of an
universal inborn distance controlling mechanism shared by man and
animals.\\
\\
{\bf Acknowledgement:} The research was supported by the Czech
Ministry of Education within the project  LC06002.  The help of the
PhD. students of the Department of Physics, University Hradec
Kralove who collected the majority of the data is gratefully
acknowledged. The help of Shinya Okazaki which was responsible for
the traffic light data is also gratefully acknowledged.

\end{document}